\documentclass[aps]{revtex4}

\usepackage{graphicx}
\usepackage{dcolumn}
\usepackage{bm}
\begin{document} 

\newcommand{\vk}{{\vec k}} 
\newcommand{{\vq}}{{\vec q}}  
\newcommand{\vx}{{\vec x}}
\newcommand{\cbar}{{\bar c}}
\newcommand{\bbar}{{\bar b}}
\newcommand{\jpsi}{{J/\psi}}
\newcommand{\av}[1]{{\langle #1 \rangle}}

\newcommand{\be}{\begin{equation}} 
\newcommand{\ee}{\end{equation}} 
\newcommand{\bea}{\begin{eqnarray}}  
\newcommand{\eea}{\end{eqnarray}} 
 
\title{Effects of Multiple Scattering in Cold Nuclear Matter \\
on $\jpsi$ Suppression and $\av{p_T^2}$ in Heavy Ion Collisions}

\author{A.M. Glenn, J.L. Nagle}
\affiliation{University of Colorado at Boulder \\ 2000 Colorado Avenue \\ 
Boulder, CO 80309, USA}
\author{Denes Molnar}
\affiliation{Department of Physics, Purdue University \\ 
525 Northwestern Avenue \\ West Lafayette, IN 47907, USA}


\begin{abstract}
Coherent multiple 
scatterings of $c\bar{c}$ quark pairs in the environment of 
heavy ion collisions have been 
used in a previous work by Qiu {\it et al.}~\cite{Qiu:1998rz} 
to study $\jpsi$ suppression. That model suggests that heavy quark 
re-scatterings in a cold nuclear medium can completely explain the 
centrality dependence of the observed $\jpsi$ suppression 
in $Pb{+}Pb$ collisions at the SPS~\cite{Abreu:2001kd}.
Their calculations also revealed significant differences 
under the assumptions of a color singlet or color octet production mechanism. 
A more recent analytic calculation~\cite{Fujii:2003fq}, which includes 
incoherent final-state re-scatterings
with explicit momentum transfer fluctuations
in {\em three} dimensions, indicates much less suppression and little 
sensitivity to
the production mechanism.
In this article, we study simultaneously both the $\jpsi$ suppression and 
$p_T$ modifications, 
at SPS and RHIC energies. We mainly focus on incoherent momentum 
transfer fluctuations 
in {\em two} 
dimensions, which is more appropriate for
the heavy-ion collision kinematics. Our analytic and Monte-Carlo calculations 
reinforce the analytic results in \cite{Fujii:2003fq}. Additionally, we find 
that the experimental $\jpsi$ suppression and
$\av{p_T^2}$ from nucleus-nucleus collisions at the SPS or RHIC cannot 
simultaneously be described in this incoherent multiple scattering framework 
for any 
value of the fluctuation strength parameter $\av{k_T^2}$.

\vspace{.2cm} 
\end{abstract} 

\maketitle 

\section{Introduction}

Heavy quarkonia ($c\cbar$, $b\bbar$)
have long been of great theoretical and experimental interest as sensitive
probes of color deconfinement and thermalization
in heavy-ion collisions~\cite{MatsuiSatz,Thews:2005vj}.
The survival probability of these bound states depends on
the density and (effective) temperature of the system, leading to the 
expectation of progressively suppressed quarkonia yields
with increasing collision centrality and/or center-of-mass energy.

\begin{figure}
\begin{center}
\vspace{-0.75cm}
\includegraphics[height=.35\textheight]{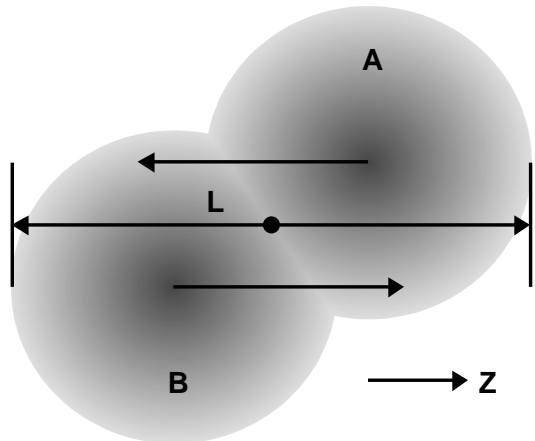}
\vspace{-0.75cm}
\caption{Diagram of a nucleus-nucleus collision. 
Arrows at the center of each nucleus indicate
the direction of travel. The dot in the center represents a 
nucleon-nucleon collision, 
and the distance $L$ depicts
the amount of cold nuclear material a product created in the nucleon-nucleon 
collision will pass through. 
The shading of the nuclei indicate the non-uniform density which should be 
accounted for when calculating $L$.}
\label{fig_collision_geo}
\end{center}
\end{figure}

The most extensively studied quarkonium state is the $\jpsi$. 
The pedagogical picture of $\jpsi$ formation in nucleus-nucleus collisions 
proceeds in
multiple stages.  
First, two Lorentz contracted nuclei pass though one another and a particular 
partonic 
hard scattering forms a $c\bar{c}$ pair, a
process requiring $t_{c\bar{c}} \sim \frac{1}{2m_c} \sim 0.01$ fm/$c$. 
The pair is then swept through the remaining fast traveling cold nuclear 
material, of 
length $L$, as shown in Figure~\ref{fig_collision_geo}.  This crossing time is 
up to $t_{cross}\sim {Diameter_{nucleus}}/{\gamma} \sim 0.1$ fm/$c$ at RHIC 
(in the center-of-mass frame). A surviving $c\bar{c}$ pair can form a 
$\jpsi$ at $t_{\jpsi}\sim Radius_{\jpsi}/c\sim 0.3$ fm/$c$ unless the hot 
dense medium left 
in the wake of the nuclear collision, lasting $t_{medium} \sim 10$ fm/$c$, 
interferes. 
With a suitable model, $\jpsi$ measurements can therefore
help extract, or at least constrain, the properties of the dense 
medium created in heavy-ion collisions.

A theoretical framework for calculating $\jpsi$ suppression has been proposed
by Qiu, Vary, and Zhang~\cite{Qiu:1998rz}.
In their approach (QVZ), $\jpsi$ production factorizes into two stages: 
first perturbative production of $c\cbar$ pairs, followed at a much
later stage by $\jpsi$ formation. 
The corresponding $A+B \to \jpsi + X$ production cross section at 
leading-order in the strong coupling is dependent on the transition 
probability, $F(q^2)$ for $c\cbar$ to evolve into a final 
$\jpsi$, where $q^{2}$ is the square of the relative momentum of the 
$c\cbar$ pair.  Various $\jpsi$ formation mechanisms can be accommodated 
via different $F$ functions. In QVZ, 
the $c\cbar \to \jpsi$ transition probabilities in
the color singlet and color octet channels were parameterized as
\bea
\label{eq_FG}
 F^{(S)}_{c\bar{c}\rightarrow \jpsi}(q^2)
 &=& N_{\jpsi}^{(S)} \, \theta(q^2)\, \exp[-q^2/(2\alpha_F^2)] 
 \qquad\qquad\qquad\qquad\qquad\qquad\qquad 
{\rm \,\,(color \,\, singlet)}\\
\label{eq_FP}
F^{(O)}_{c\bar{c}\rightarrow \jpsi}(q^2)
&=& N_{\jpsi}^{(O)} \, \theta(q^2)\, \theta(4m'^2-4m^2_c-q^2)\times
\left(1-\frac{q^2}{4m'^2-4m^2_c}\right)^{\alpha_F}
\qquad {\rm \,\,(color \,\, octet)}
\eea
where $m_c$ is the charm quark mass, $m'$ 
is the mass scale for the open charm threshold ($D$ meson mass), 
and $N_{\jpsi}$ and $\alpha_F$ are parameters fixed from hadron-hadron 
collision data.

In a nuclear environment, QVZ includes $\jpsi$ suppression via
parton multiple scatterings in cold nuclear matter 
for the $c\cbar$ pair only, and does not consider further suppression
in the hot medium created later.
Multiple scattering of the $c\cbar$ pair increases the $q^2$ of the pair, 
reducing the overlap with 
$F(q^2)$. 
The original QVZ modeled this effect 
with a {\em constant} $q^2$ shift linear in nuclear 
pathlength~\cite{Benesh:1994,Qiu:1998rz}
\bea
\label{eq_lin}
q^2 \to q^2 + \Delta q^2 = q^2 + \varepsilon^2 L \ ,
\eea
which was an intuitive generalization of an identical result for
the nuclear broadening in photoproduction of jets given by twist-4
contributions~\cite{Luo:1993ui}.
Surprisingly, this 
approach was able to reproduce not only hadron-nucleus ($h{+}A$) but also
nucleus-nucleus ($A{+}A$) data up to SPS energies, 
including the ``anomalous'' suppression in $Pb{+}Pb$~\cite{Abreu:2001kd},
with the {\em same} $\varepsilon^2\sim 0.2-0.3$~GeV$^2$/fm. 
The provocative result indicated negligible 
additional $\jpsi$ suppression from the hot medium in energetic heavy-ion
collisions, contrary to expected signatures of a dense parton plasma.



A more detailed analysis of coherent multiple scattering effects in 
the Drell-Yan process~\cite{Fries:2003} also found a
constant $\Delta q^2 \propto L$, provided one
considers at each twist only the contribution that gives the 
largest $q^2$ change. The neglected terms, 
on the other hand, 
would affect the lower $\Delta q^2$ region and therefore
generate fluctuations in $\Delta q^2$ at fixed $L$.
Here we study fluctuations in 
multiple scatterings of charm quarks and antiquarks
in the opposite, {\em incoherent} scattering limit.
In that case, 
fluctuations in $\Delta q^2$  
are of the same order of magnitude as the average $\Delta q^2$
for the $c\cbar$ pair. An earlier study\cite{Fujii:2003fq}, which 
considered momentum transfers in three dimensions,
found that fluctuations in $\Delta q^2$ lead to
much weaker suppression in $A{+}A$ than in the QVZ approach.
While such multiple scatterings of the heavy quarks are expected to
have a 
direct impact on the $\jpsi$ $p_T$ 
distributions~\cite{Kharzeev:1997ry,Nagle:1999ms},
the QVZ and Fujii~\cite{Fujii:2003fq} calculations do not provide any 
information about $p_T$. In this work we focus on the interplay between 
the $\jpsi$ yield and $\av{p_T^2}$ based on a Monte-Carlo approach.

\section{Role of momentum transfer fluctuations}
\label{Sec_fluct}

In the QVZ framework, 
final-state quark scatterings affect the momentum distribution of $c\cbar$ 
pairs and therefore the $\jpsi$ yield as 
\be
dN_{\jpsi} = F(q^2)
\int d\Delta q^2 P(\Delta q^2;q^2-\Delta q^2) \, 
dN_{c\cbar}(q^2-\Delta q^2) \equiv
F(q^2) \, d\bar N_{c\cbar}(q^2)  \ ,
\ee
where $P(\Delta q^2; q^2)$ 
is the probability of accumulating a total $\Delta q^2$ change in
the relative pair momentum given an initial value $q^2$.
A completely equivalent approach is to group 
scattering effects into a modified formation probability $\bar F$
\be
dN_{\jpsi} = dN_{c\cbar}(q^2) 
\int d\Delta q^2 P(\Delta q^2; q^2) \, 
F(q^2+\Delta q^2) \equiv \bar F(q^2) 
\, dN_{c\cbar}(q^2) \ ,
\label{Feff}
\ee
which has the advantage that
the primary $c\cbar$ distribution and final state effects factorize.

The effective formation probability can be calculated from a model of
final state interactions in the medium.
We consider here incoherent 
Gaussian momentum kicks in {\em two} dimensions 
transverse to the beam axis, independently 
 for each quark (i.e., ignore correlations between $c$ and $\cbar$)%
, as a model of small-angle scatterings off fast partons. This
results in momentum-space
random walk probability distributions
\be 
g(\vk_{T,i}) = \frac{2}{\pi \, L\,  \varepsilon^2} 
             \exp\!\!\left[-\frac{2\vk_{T,i}^{\ 2}}{L\, \varepsilon^2}\right] 
\qquad (i=1,2 {\rm \ for\ } c,\cbar) \ ,
\ee
where $\vk_{T,i}$ is the total momentum transferred, while
$\varepsilon^2/2$ is average 
momentum-squared transferred per unit nuclear pathlength.
The average momentum and the average relative momentum transferred 
to the $c\cbar$ {\em pair}, $\vk_T \equiv \vk_{T,1} + \vk_{T,2}$ and
$\Delta \vk_T \equiv \vk_{T,1} - \vk_{T,2}$, 
are also Gaussian but with 
dispersions $L\, \varepsilon^2$.

In the nonrelativistic limit, the accumulated change in 
the relative pair momentum is
\be
\Delta q^2 \approx \Delta \vq^{\ 2} = 
(\vq + \vk_T)^2 - \vq^{\ 2} 
= \Delta k_T^2 + 2\Delta k_T q \cos \theta \ .
\label{eq_dq2}
\ee
Final state scatterings can be considered at
three levels of sophistication:

\medskip
\noindent
{\em i) constant shift (CS)} 
- keep only the {\em average} of the first term, i.e.,
$\Delta q^2 \to L\, \varepsilon^2$ as in (\ref{eq_lin});

\medskip
\noindent
{\em ii) partial fluctuations (PF)} -
keep fluctuations in the magnitude $\Delta k_T^2$ but ignore the 
second angular term, i.e., $\Delta q^2 \to \Delta k_T^2$; and

\medskip
\noindent
{\em iii) all fluctuations (ALL)} - keep
fluctuations both in magnitude and direction, i.e., the full expression 
(\ref{eq_dq2}).

\medskip
\noindent
In the color singlet case, the effective transition probabilities are 
calculable analytically in a straightforward manner:
\bea
\bar F^{(S)}_{CS}(q^2)
&=&N_{\jpsi}^{(S)} 
\exp\!\!\left[-\frac{q^2+L\,\varepsilon^2}{2\alpha_F^2}\right]
\nonumber \\
\bar F^{(S)}_{PF}(q^2) 
&=&N_{\jpsi}^{(S)} \frac{2\alpha_F^2}{2\alpha_F^2+L\,\varepsilon^2}
\exp\!\!\left[-\frac{q^2}{2\alpha_F^2}\right]
\nonumber \\
\bar F^{(S)}_{ALL,2D}(q^2) 
&=& N_{\jpsi}^{(S)} \frac{2\alpha_F^2}{2\alpha_F^2+L\,\varepsilon^2}
\exp\!\!\left[-\frac{\vq_T^{\ 2}}{2\alpha_F^2+L\,\varepsilon^2} - 
\frac{Q_0^2}{2\alpha_F^2}\right]
\nonumber \\
&=& 
N_{\jpsi}^{(S)} \frac{2\alpha_F^2}{2\alpha_F^2+L\,\varepsilon^2}
\exp\!\!\left[-\frac{q^2}{2\alpha_F^2+L\,\varepsilon^2}\right] \,
\exp\!\!\left[- \frac{L\varepsilon^2 Q_0^2}{2\alpha_F^2 
(2\alpha_F^2 +L\varepsilon^2)}\right] 
\ 
\nonumber \\
\bar F^{(S)}_{ALL,3D}(q^2)
&=& 
N_{\jpsi}^{(S)} 
\left(\frac{3\alpha_F^2}{3\alpha_F^2+L\,\varepsilon^2}\right)^{3/2}
\exp\!\!\left[-\frac{3\vq^{\ 2}}{6\alpha_F^2+2L\,\varepsilon^2} + 
\frac{q_0^2}{2\alpha_F^2}\right]
\nonumber \\
&=& 
N_{\jpsi}^{(S)} 
\left(\frac{3\alpha_F^2}{3\alpha_F^2+L\,\varepsilon^2}\right)^{3/2}
\exp\!\!\left[-\frac{3q^2}{6\alpha_F^2+2L\,\varepsilon^2}\right] \,
\exp\!\!\left[\frac{L\varepsilon^2 q_0^2}{2\alpha_F^2 
(3\alpha_F^2 +L\varepsilon^2)}\right] 
\ ,
\label{Feff_S}
\eea
where $Q_0^2 \equiv q_z^2- q_0^2$, 
$q^2 \equiv \vq_T^{\ 2} + Q_0^2$, and the last two lines are for Gaussian 
scattering in three dimensions. 
If fluctuations are neglected, $\bar F$ is strongly suppressed, 
exponentially with pathlength. 
This results in exponential $\jpsi$ suppression with a suppression 
factor $R_{\jpsi} = \exp[-L\,\varepsilon^2/(2\alpha_F^2)].$
Fluctuations in the magnitude of $\Delta \vk_T$
change the behavior to a milder power-law reduction with $L$, 
$R_{\jpsi} = 2\alpha_F^2 / (2\alpha_F^2+L\,\varepsilon^2)$.
The biggest effect, however, comes from 
fluctuations in the direction of $\Delta \vk_T$. 
These broaden $\bar F$ compared
to the original $F(q^2)$, further weakening suppression effects.
It is easy to understand why angle fluctuations are important. 
From any starting point 
in momentum space,
there is always a finite probability for the random walk to get {\em closer}
to the origin, thereby enhancing the formation probability compared
to the other two approximations where any initial $q^2$ can only grow.

\begin{figure}
\begin{center}
\includegraphics[height=6cm]{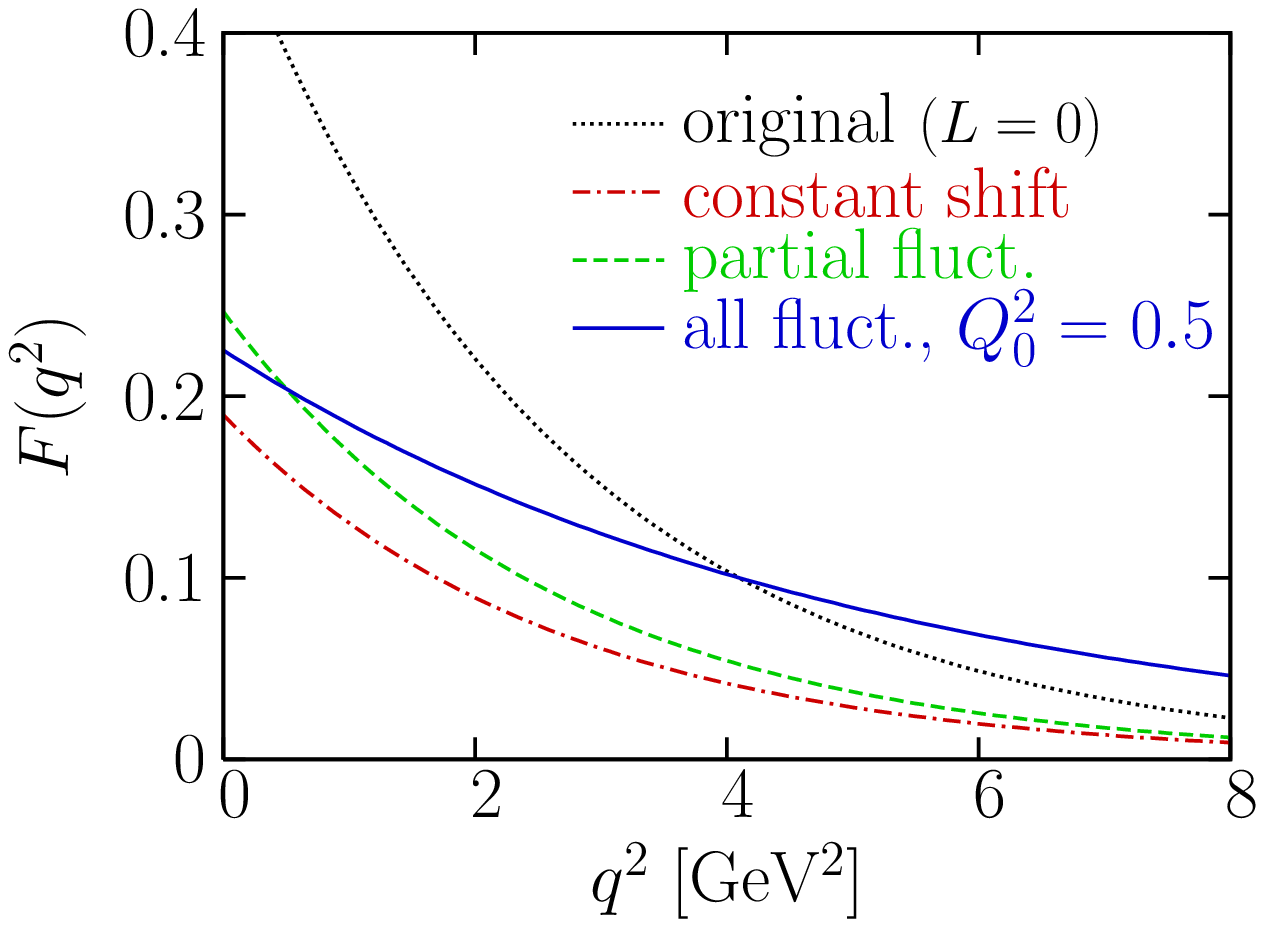}
\includegraphics[height=6cm]{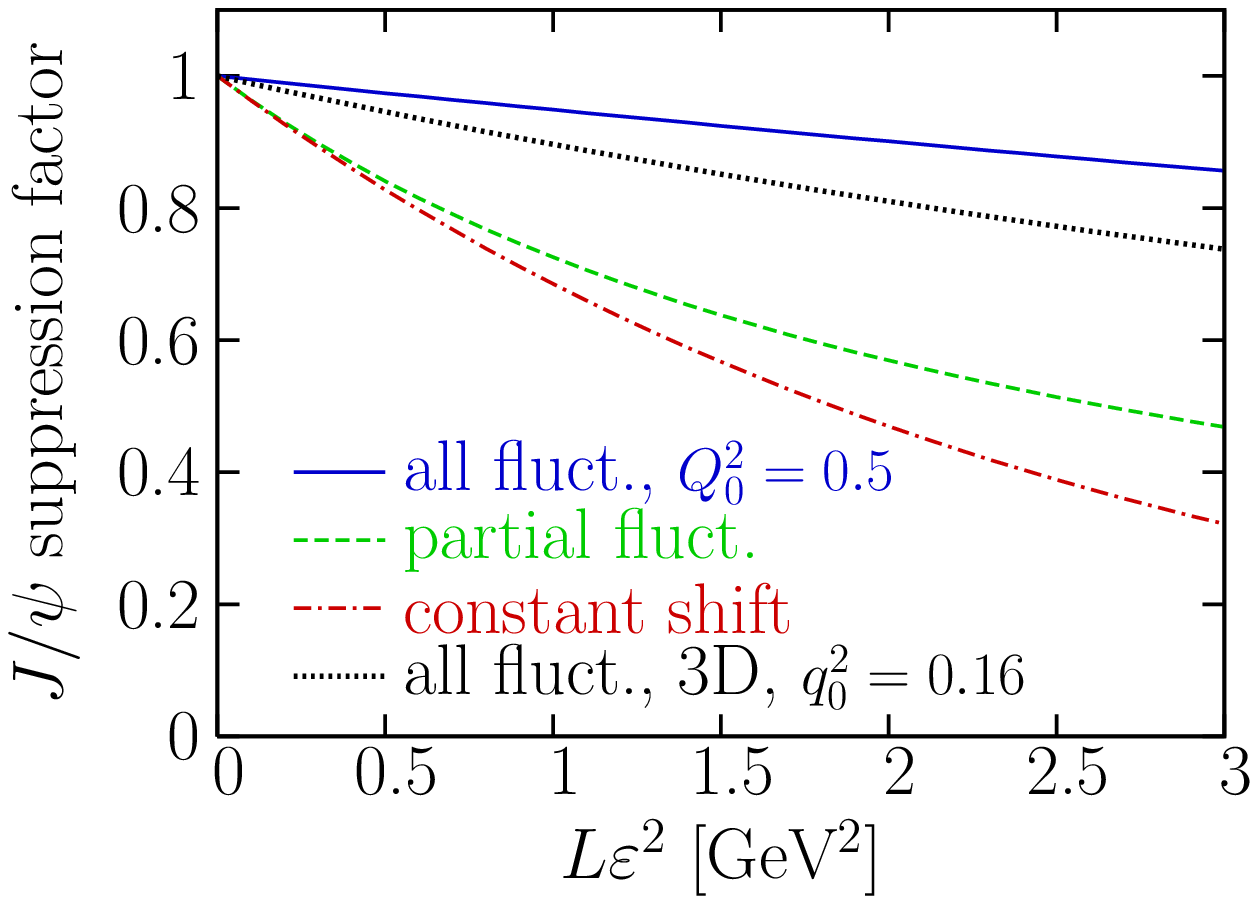}
\vspace{-0cm}
\caption{{\em Left plot:} effective $c\cbar\to \jpsi$ transition probability 
in the color-singlet channel for $L\,\varepsilon^2 = 8\times 0.3$~GeV$^2$ 
(characteristic of near-central $Pb+Pb$ or $Au+Au$ collisions) with
three different approaches to treat fluctuations in final-state interactions.
Results for the constant shift (dashed-dotted)
and ``partial fluctuation'' (dashed) approximations (see text)
are compared to the full 2D Gaussian random walk result, assuming 
$Q_0^2= 0.5$~GeV$^2$, (solid) 
and the case of no final-state
interactions (dotted). The full result is much broader in $q^2$,
resulting in much weaker $\jpsi$ suppression.
{\em Right plot:} suppression of the total $\jpsi$ in nuclear collisions at
$\sqrt{s_{NN}}=200$~GeV as a function of 
$L\,\varepsilon^2$ for the constant shift (dashed-dotted), partial fluctuation 
(dashed) approximations, and the full 2D Gaussian random walk result 
(solid).
The full result gives much smaller suppression than the two approximations.
Results for the 3D random walk in~\cite{Fujii:2003fq} are also shown (dotted), 
for $q_0^2 = 0.16$~GeV$^2$.
}
\label{fig_fluct}
\end{center}
\end{figure}

Figure~\ref{fig_fluct} illustrates the influence of fluctuations
and confirms the above general considerations.
The left plot shows the color singlet $\bar F(q^2)$ 
resulting from the different fluctuation treatments, for
the original QVZ parameter set $N_{\jpsi} = 0.47$, 
$\alpha_F = 1.15$, and with $\varepsilon^2 = 0.3$~GeV$^2/$fm 
and a pathlength $L=8$~fm characteristic of
near-central $Pb+Pb$ or $Au+Au$ collisions.
The right plot shows the corresponding 
$\jpsi$ suppression factors (relative to the 
$L=0$ case) as a function of $L\,\varepsilon^2$, for an initial $c\cbar$ 
distribution in $p+p$ events at $\sqrt{s} = 200$~GeV 
from the PYTHIA~\cite{pythia_params} event generator program
(the PYTHIA results can be parameterized as
$dN_{c\cbar}/dq^2 \propto (q^2)^{0.518}
 / [2.08\cdot 10^6+(\sqrt{q^2/(1{\rm \ GeV^2})}+4.04)^{8.15}]$).
Clearly, the largest suppression comes from the simplest,
 constant shift approximation (dashed-dotted line). 
On the other hand, the inclusion of 
fluctuations gives a much-reduced suppression (solid line).
For comparison we also show the result for the {\em three-dimensional}
Gaussian random walk
considered in \cite{Fujii:2003fq} (dotted line), which gives
a stronger suppression than the more realistic 2D case. 
The reason is that in 3D,
angle fluctuations have a weaker role because the solid angle 
$\sin\theta d\theta d\phi$ 
prefers $\theta$ values near $\pi/2$, for which 
the angular term in (\ref{eq_dq2}) is small.
Though in the very-large-$L$ limit the suppression factor behaves as $1/L$ and
$1/L^{3/2}$ for 2D and 3D random walk respectively~\cite{Fujii:2003fq},
we found that in practice these limits apply poorly and 
give more suppression at moderate $L\,\varepsilon^2 \sim  0.5-5$~GeV$^2$ 
than even the constant shift approximation.
All the above trends apply to the color octet case as well.

We note that in the simple analytic calculation above
we applied (\ref{Feff_S})
with a constant $Q_0^2 = 0.5$~GeV$^2$ (2D case) 
and $q_0^2 = 0.16$~GeV$^2$ (3D case), which are the average values
for midrapidity $\jpsi$-s coming from 
the region $q^2 \sim 0-3$~GeV$^2$
that gives the dominant contribution to the yield~\cite{denes:foot2}. 
In general, the various $q$ components
are correlated (e.g., $Q_0^2 \le q^2$) and therefore the calculation
would require $dN/dq^2 dQ_0^2$ (2D) or $dN/dq^2 dq_0^2$ (3D)
in analytic form as an input. 
We include the $q^2-Q_0^2$ correlations for 2D random walk
in Sec.~\ref{MC} using a Monte-Carlo approach.

If the strength of momentum kicks per unit pathlength $\varepsilon^2$ were
an unknown parameter, much of the above discussion would be academic
because we found that in the calculation above 
each approximation can reproduce the exact result 
quite well
with an appropriate rescaling of $\varepsilon^2$. 
In particular,
$\varepsilon^2_{CS} : \varepsilon^2_{PF} : \varepsilon^2_{ALL,2D} 
: \varepsilon^2_{ALL,3D} \approx 1 : 1.5 
: 9.5 : 5$~\cite{denes:footnote}.
However, an independent determination of $\varepsilon^2$
from FermiLab data on dijet momentum imbalance already
constrains $\varepsilon^2$ to $\sim 0.2-0.5$~GeV$^2$/fm~\cite{Luo:1993ui}.

Even if $\varepsilon^2$ were arbitrary, 
it affects not only the $\jpsi$ yields but also the spectra. 
In particular the larger the $\varepsilon^2$, the higher the  
$\av{p_T^2}$ of the $\jpsi$'s.
In the following sections
we analyze the suppression-$\av{p_T^2}$ consistency 
in detail and contrast it to data from the SPS and RHIC,
using a Monte-Carlo 
approach.
For the spectra, both initial and final state scatterings are important.

\section{Monte-Carlo Study}
\label{MC}

In this study we utilize a Monte-Carlo Glauber model~\cite{Glauber:1970jm} 
calculation, 
as implemented in the Heavy Ion community~\cite{Back:2001xy,Adcox:2001jp},
which assumes a 30~mb and 42~mb nucleon-nucleon 
cross section for $\sqrt{s_{NN}} = 17.2$ and 200~GeV respectively, 
for collision 
geometry.
We consider a nucleus-nucleus collision at a given impact parameter
and collision energy, 
and for each binary collision assign a $c\bar{c}$ pair with momentum 
vectors from charm events in $p{+}p$ 
collisions (at the same $\sqrt{s_{NN}}$) 
from PYTHIA 6.205 with the parameters used in 
Section~\ref{Sec_fluct}~\cite{pythia_params}.
The pathlength depicted in Fig.~\ref{fig_collision_geo},
$L \equiv L_{A} + L_{B}$, is the
sum of contributions by nucleons
moving towards the production point from the left and right,
which are calculated 
by integrating over the nuclear density distribution 
and dividing by
the density at the center of the nucleus.
E.g., for a production point $(\vx_T,z_0)$ relative to the center of 
nucleus A 
\be
L_{A}(\vx_T,z_0) 
\equiv 
\frac{1}{\rho_{max}} \int_{z_0}^{\infty} dz\, 
\rho(\sqrt{z^2 + \vx_T^{\,2}})
= \frac{\av{N_{coll}(\vx_T,z_0)}}{\rho_{max}\,\sigma_{inel}^{N+N}} \ ,
\label{def_L}
\ee
where $\av{N_{coll}}$ is the average
number of remaining binary collisions on the
right and $\sigma_{inel}^{N+N}$ is the inelastic nucleon-nucleon cross section.
Final state scatterings are modeled via a 2D Gaussian random walk as 
described in Section \ref{Sec_fluct} except that, instead of 
the continuous $L \varepsilon^2$ variable, 
the number of scatterings is quantized in terms of the binary collisions,
where the average momentum transfer squared per scattering is 
$\av{k_T^2}$; i.e., in the Monte-Carlo $N_{coll} \av{k_T^2}$ is equivalent 
to $L\, \varepsilon^2$ in the analytic calculations. 
We confirmed via increasing 
the kicks per binary collision, while reducing the $\av{k_T^2}$ by 
the same factor, 
that the discretization does not have a significant effect
on our results.
For initial state scatterings, which are meant to represent the scattering of 
individual 
incoming partons (mainly gluons) before the $c\cbar$ production, the $k_T$ is 
transfered to
the $p_T$ of the $c\cbar$ pair with no increase in the $q^2$.  
For simplicity, we assume that the $\av{k_T^2}$ for initial state 
parton scatterings is the same as for the final state parton scatterings.

From Eq.~(\ref{def_L}) it is clear that the $\av{k_T^2}$ 
parameter in the Monte-Carlo is proportional to
$\varepsilon^2$, the average increase in $q^2$ per unit 
nuclear pathlength.
The constant of proportionality depends on the collision energy
(through the inelastic nucleon-nucleon
cross section) and also the nuclear density at the center of the 
colliding nuclei.
For convenience, we tabulate in Table~\ref{table_epsilon}
several $\av{k_T^2}$ values and the corresponding $\varepsilon^2$,
for $Pb+Pb$ collisions at $\sqrt{s_{NN}} = 17.2$~GeV (top SPS energy). 

\begin{figure}
\begin{center}
\includegraphics[height=.4\textheight]{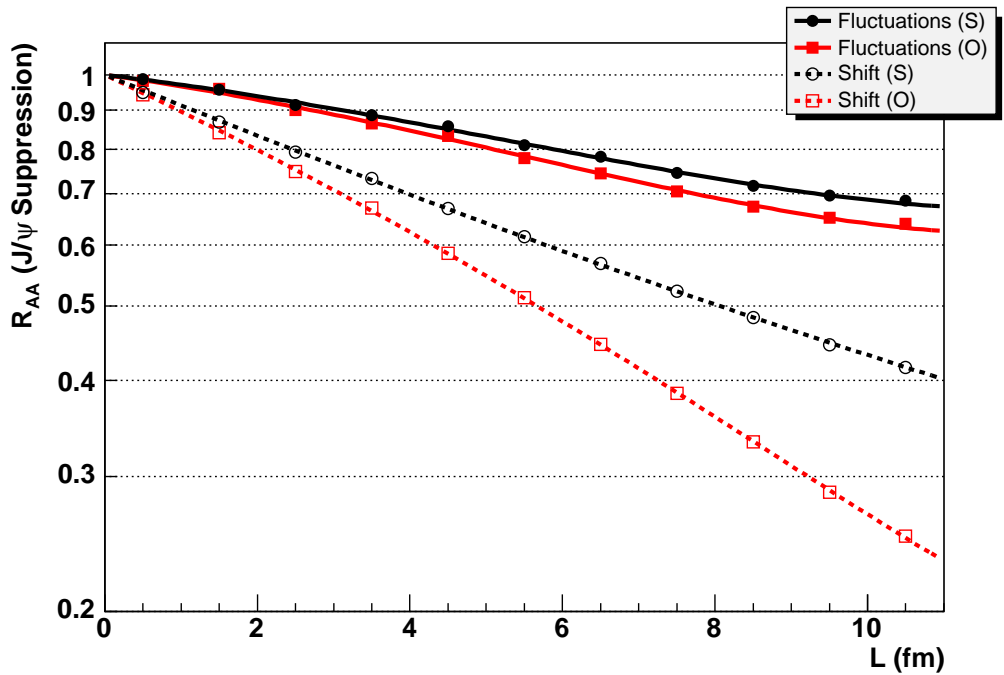}
\caption{The dependence of $\jpsi$ suppression on collision centrality for 
$\sqrt{s_{NN}}=17.2$~GeV
$Pb{+}Pb$ collisions for $\av{k_T^2}=0.3$~GeV$^2$.
The filled markers 
indicate results from the model with $q^2$ 
changes due to re-scattering including fluctuations. The open markers show 
results when the explicit re-scattering is replaced by an
overall shift which results in the same 
average $q^2$ increase per unit pathlength,
$\Delta q^2 /L \equiv \varepsilon^2=0.27$~GeV$^2$/fm.
``S''(circles) and ``O''(squares) refer to the color singlet and 
color octet transition probabilities 
(see Eqs. (\ref{eq_FG}) and (\ref{eq_FP})).} 
\label{fig_shift_vs_fluctuation}
\end{center}
\end{figure}

As a consistency check, instead of our multiple scattering model, 
we study $\jpsi$ suppression as a function of $L$ for
$\sqrt{s_{NN}}=17.2$~GeV $Pb{+}Pb$ collisions
via the $\bar{q}^2=q^2+\varepsilon^2L$ shift as done by QVZ.
The only difference is that we use the PYTHIA $q^2$ distribution 
for $c\bar{c}$ pairs.
The results, shown in Fig.~\ref{fig_shift_vs_fluctuation} 
and labeled 'Shift', 
give similar suppression trends to those reported in~\cite{Qiu:1998rz}.
In particular, we observe the larger suppression 
as a function of $L$ for the color octet transition probability 
$F^{(O)}(q^2)$ (defined in Eq.~(\ref{eq_FP})).  For the Monte-Carlo 
studies, we use 
$N_{\jpsi}=0.47$ and $\alpha_F=1.2$ for $F^{(S)}(q^2)$ and 
$N_{\jpsi}=0.485$ and $\alpha_F=1$ for $F^{(O)}(q^2)$ (consistent with 
Sec. \ref{Sec_fluct} and QVZ~\cite{Qiu:1998rz}).

\begin{table}
\begin{center}
\caption{The $\varepsilon^2 \equiv \Delta\av{q^2}/L$ values 
corresponding to various $\av{k_T^2}$ parameter values for 
$\sqrt{s_{NN}}=17.2$~GeV $Pb+Pb$ collisions.}
\begin{tabular}{ccc}
\hline
 \raisebox{-0.05cm}{$\av{k_T^2}$}  & 
  \raisebox{-0.05cm}{$\varepsilon^2 \equiv \Delta\av{q^2}/L$}\\
(GeV$^2$)  & (GeV$^2$/fm)\\
\hline
0.1 & 0.09\\
0.2 & 0.18\\
0.3 & 0.27\\
0.4 & 0.36\\
\hline
\label{table_epsilon}
\end{tabular}
\end{center}
\end{table}

Figure~\ref{fig_shift_vs_fluctuation}
also shows a comparison of $\jpsi$ suppression
in the shifted $q^2$ model and the fluctuation model.
Just like for the analytic study in Section~\ref{Sec_fluct},
we keep the average $q^2$ shift the same in both cases and therefore
use $\av{k_T^2} =0.3$ 
GeV$^2$ in the Monte Carlo, which  corresponds to
$\varepsilon^2 = 0.27$~GeV$^2/$fm,
a shift consistent with range used by QVZ in \cite{Qiu:1998rz}.
We find that fluctuations give
significantly less suppression.
Clearly,
an incoherent random walk type process cannot be 
consistent with the $\bar{q}^2=q^2+\varepsilon^2L$ shift since 
the most probable scattering value 
is zero and the low $q^2$ region therefore cannot be completely depleted.  
We note that our numerical results agree with 
the analytic ones presented for the singlet case in Section \ref{Sec_fluct}.

In addition, the suppression patterns for 
the color singlet and color octet transition probabilities, 
$F^{(S)}(q^2)$ and $F^{(O)}(q^2)$, become nearly indistinguishable
when fluctuations are included.
Therefore, we only report results for the octet case
for the remainder of this study.
This is in agreement with the previous analytic 
calculations~\cite{Fujii:2003fq} that
incorporated momentum transfer fluctuations in three dimensions.

\section{Results}

\begin{figure}
\begin{center}
\includegraphics[height=.36\textheight]{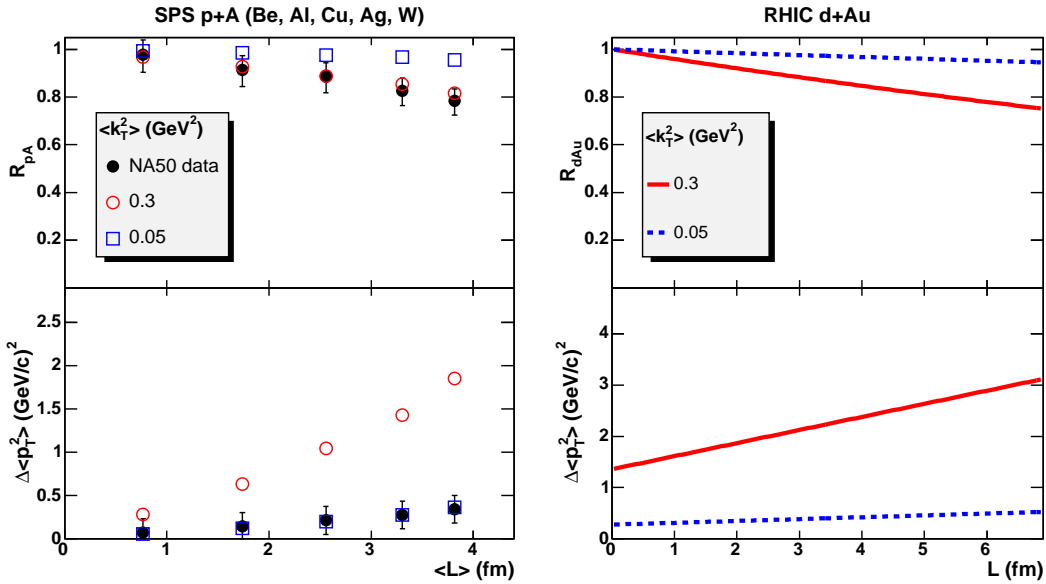}
\caption{{\em Left plots:} The nuclear modification factor (Upper) and 
$\Delta\av{p_T^2}$ (Lower) for $\jpsi$ for 400~GeV/$c$ 
proton beams incident on several fixed targets 
as a function of the mean nuclear path-length $\av{L}$. 
The filled circles show data from NA50 measurements~\cite{NA50}. 
The suppression data uses the $p{+}p$ value 
scaled from 450~GeV/$c$, and the $p{+}p$ reference value for $\av{p_T^2}$
of $1.6$~(GeV/$c)^2\pm10\%$ comes from the projection of 
a fit to the $\av{p_T^2}$ $p{+}A$ data to $L=0$~fm. The errors for both
graphs are dominated by the $p{+}p$ reference, so they are highly correlated. 
{\em Right plots:} calculations for $\sqrt{s_{NN}}=200$~GeV $d{+}Au$ collisions.
Note that for the left plots each $\av{L}$ comes from a different target nucleus, while for
the right plots the $L$ is for different production point geometries. 
The calculations use the $F^{(O)}(q^2)$ transition probability.
}  
\label{fig_pAdAcomp}
\end{center}
\end{figure}

We first use the model to study suppression and $p_T$ distributions
in $p(d){+}Au$ collisions at SPS and RHIC energies. High statistics data
are available for comparison from the NA50 experiment. In 
Figure~\ref{fig_pAdAcomp} 
we show results for $\av{k_T^2}$ values which reasonably reproduce either the 
$\Delta\av{p_T^2} = \av{p_T^2}_{p{+}A} - \av{p_T^2}_{p{+}p}$ or the 
suppression as a function of the nuclear path-length. 
For $p{+}A$, a $\av{k_T^2}=0.3$~GeV$^2$ reproduces the 
$R_{pA}$ dependence, but produces far too much $\Delta\av{p_T^2}$ increase. A 
$\av{k_T^2}=0.05$~GeV$^2$, which is consistent with the $\Delta\av{p_T^2}$ trend, 
under predicts the suppression for the largest nuclei.
Even for $p{+}A$ the model cannot 
simultaneously give satisfactory descriptions of $\Delta\av{p_T^2}$ and 
$R_{pA}$ for any $\av{k_T^2}$ for large nuclei. 
We note that the relative strength of initial and final state scatterings is 
not tuned in our model, but since the initial state scatterings currently account for
roughly half of the $\Delta\av{p_T^2}$ increase, removing initial 
state scatterings would not be sufficient to bring the 
model into good agreement with the data.
Calculations for $\sqrt{s_{NN}}=200$~GeV $d{+}Au$ using the 
same $\av{k_T^2}$ values are also shown in Figure \ref{fig_pAdAcomp}.

Experimental measurements of $\jpsi$ suppression $R_{dAu}$ and
$\Delta\av{p_T^2}$ have recently been 
made by the PHENIX collaboration at RHIC~\cite{Adler:2005ph}.  
They report 
$\av{p_T^2} _{d{+}Au}
-\av{p_T^2}_{p{+}p} = 1.77\pm0.37$,
$1.12\pm0.33$, and $-1.28\pm0.94$ and $R_{dAu} = 1.18\pm0.12$, $0.79\pm0.06$, 
and $0.95\pm0.1$ for 
backward, forward, and mid rapidity respectively in minimum bias deuteron-gold
reactions.
Table~\ref{table_dAu} shows our Monte-Carlo results for various $\av{k_T^2}$ 
values.
Although no strong statement of consistency can be made due to current 
experimental errors, the results from our 
model for $\av{k_T^2} > 0.3$~GeV$^2$ can be safely excluded from 
simultaneously satisfying both 
experimental $R_{dAu}$ and $\Delta\av{p_T^2}$ constraints.  
We note that this level of suppression is consistent with that determined from 
a simple nuclear absorption Glauber model with a cross section 
$\sigma_{\jpsi-N} \approx 1-3$ mb~\cite{Vogt:2005ia}, 
which we have also confirmed.
Future measurements of $\jpsi$ in deuteron or proton-nucleus reactions with 
sufficient
statistics for centrality dependencies to be accurately determined will be
important to narrow this parameter range.

\begin{table}
\begin{center}
\caption{Results for $\jpsi$ properties from the model for minimum bias 
collisions 
in $\sqrt{s_{NN}}=200$~GeV $d{+}Au$ for several values of input $\av{k_T^2}$.}
\begin{tabular}{ccc}
\hline
$\av{k_T^2} $ & Suppression & $\Delta\av{p_T^2}$ \\
(GeV$^2$) & ($R_{dAu}$) &  (GeV$^2$) \\
\hline
0.0 & 1.00 &  0.0 \\
0.075 & 0.95 &  0.6 \\
0.15 & 0.91 &  1.3 \\
0.225 & 0.87 &  1.8 \\
0.3 & 0.84 &  2.4 \\
\hline
\label{table_dAu}
\end{tabular}
\end{center}
\end{table}

\begin{figure}
\begin{center}
\includegraphics[height=.4\textheight]{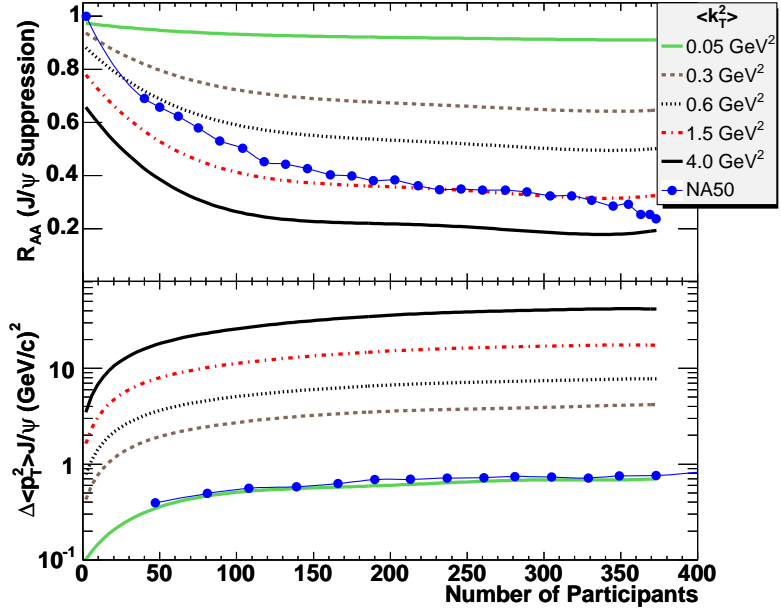}
\caption{The nuclear modification factor and 
$\Delta\av{p_T^2}$ for $\jpsi$ in $\sqrt{s_{NN}}=17.2$~GeV 
$Pb{+}Pb$ collisions as a function of the number of 
participants for several input values of $\av{k_T^2}$.  
The filled circles show 
data from NA50 
measurements~\cite{Abreu:2000xe,Abreu:2001kd}. The calculations use the 
$F^{(O)}(q^2)$ transition probability.} 
\label{fig_NA50comp}
\end{center}
\end{figure}

Despite the inability of this model to fully describe the $p(d){+}Au$ 
data, we feel it is useful to follow through with heavy ion results. 
For a high-temperature quark-gluon plasma,
the incoherent final-state scattering assumption is more
justified because the Debye length is much smaller than the mean free path 
$\mu_D^{-1} \sim (gT)^{-1} \ll \lambda \sim (g^2 T)^{-1}$~\cite{denes:footnote3}. 
Here we report results 
for $Pb{+}Pb$ collisions at $\sqrt{s_{NN}}=17.2$~GeV. 
Figure~\ref{fig_NA50comp} shows the $\jpsi$ suppression relative to binary 
collision scaling $R_{AA}$ 
and $\Delta\av{p_T^2}$ as a function of centrality for various 
$\av{k_T^2}$ values.  As $\av{k_T^2}$ is increased, the fraction 
of surviving $\jpsi$
decreases and the $\av{p_T^2}$ of the surviving $\jpsi$ 
increases.
Figure~\ref{fig_NA50comp} also shows data points from the
NA50 experiment~\cite{Abreu:2000xe,Abreu:2001kd}.
The general suppression trend is not well reproduced by any value 
of $\av{k_T^2}$.  
A value of $\av{k_T^2}$ which reproduces the suppression for the most central 
events
substantially over-predicts the increase in $\av{p_T^2}$.
The effect of heavy quark re-scattering in a cold nuclear medium, as described
in this model, cannot explain the NA50 data. 

\begin{figure}
\begin{center}
\includegraphics[height=.4\textheight]{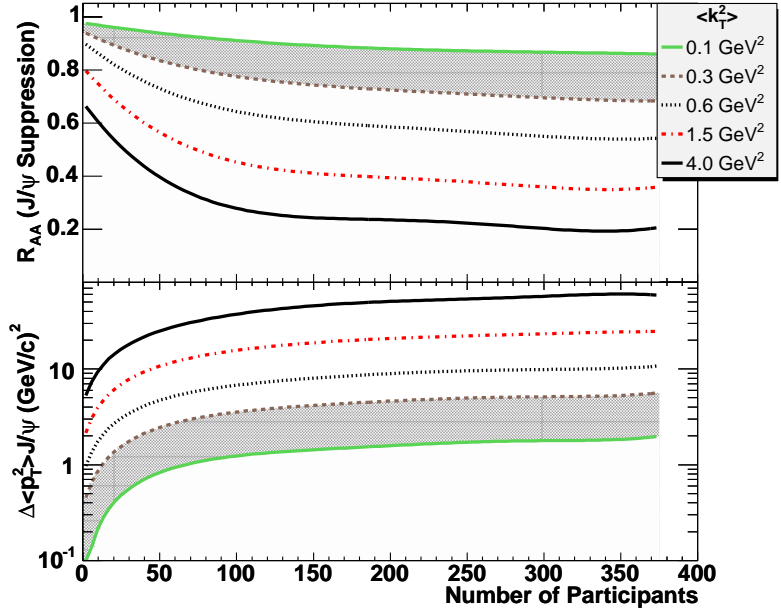}
\caption{
The nuclear modification factor and 
$\Delta\av{p_T^2}$
for $\jpsi$ in $\sqrt{s_{NN}}=200$~GeV $Au{+}Au$ collisions as a function of 
the number of participants 
for several input values of $\av{k_T^2}$.  The shaded area represents the range
 of $\av{k_T^2}$
values which are in rough agreement with the PHENIX $d{+}Au$ 
data~\cite{Adler:2005ph}. Our calculations use the color octet 
$F^{(O)}(q^2)$ transition probability.
}  
\label{fig_AuAu_pTshift}
\end{center}
\end{figure}

Lastly, we use the model to study suppression and $p_T$ distributions
in $Au{+}Au$ collisions at $\sqrt{s_{NN}}=200$~GeV.
The $R_{AA}$ and $\Delta \av{p_T^2}$ results are shown 
for $Au{+}Au$ in 
Fig.~\ref{fig_AuAu_pTshift}, which again demonstrates the strong correlation 
between 
increased suppression, from $q^2$ change,
and increased $\Delta\av{p_T^2}$ in models using this 
type of 
multiple scatterings of $c$ and $\bar{c}$ quarks to account for $\jpsi$ 
suppression. 
For $\av{k_T^2} = 0.3$~GeV$^2$, roughly the largest value consistent with 
$d{+}Au$ 
data, 
the fairly week suppression, $R_{AA}\sim0.7$, requires 
$\Delta\av{p_T^2}\sim5$~GeV$^2$. 

\begin{figure}
\begin{center}
\includegraphics[height=.4\textheight]{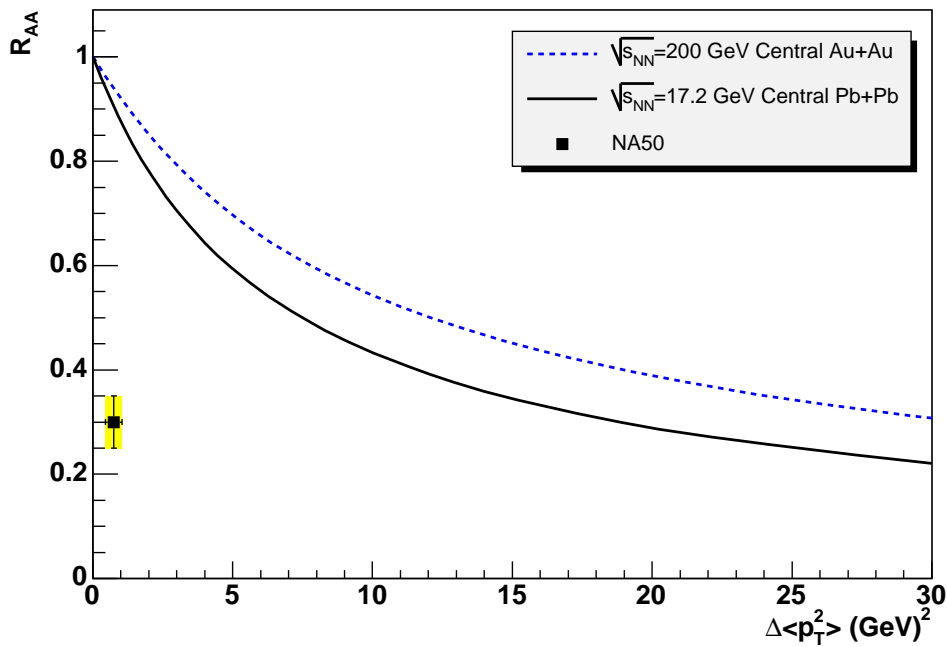}
\caption{The $R_{AA}$ vs $\Delta\av{p_T^2}$ curves 
traced by varying 
$\av{k_T^2}$ for events with 300-350 participants (central) for $Au{+}Au$ and 
$Pb{+}Pb$ 
collisions at RHIC and SPS energies respectively.  The black square shows the 
value from 
NA50 measurements~\cite{Abreu:2000xe,Abreu:2001kd}.}
\label{fig_pt2_vs_Raa}
\end{center}
\end{figure}

We highlight this connectedness and the inconsistency with 
available RHIC and SPS data for central events in Fig.~\ref{fig_pt2_vs_Raa}. 
We show that for any parameter values of $\av{k_T^2}$ one is constrained 
to be on a curve correlating suppression and $p_T$.
The SPS data are far from the $R_{AA}$ vs 
$\Delta\av{p_T^2}$ curve.
We note that preliminary data from the PHENIX
experiment also appear to be inconsistent 
with any points on the curve~\cite{hugo}.
For a given $\av{p_T^2}$, the suppression at the SPS is stronger because 
for a lower collision energy the initial $c\cbar$ $q^2$ distribution is 
narrower, and therefore the yield is more sensitive to rescattering 
(i.e., shifts in $q^2$).
Although this study does not account for nuclear modifications of the 
parton distribution function (i.e., effects such as nuclear shadowing), 
these effects do not appear significant 
enough to change the basic conclusions of our 
study~\cite{Vogt:2005ia,Vogt:2004dh}.

\section{Conclusions}

We have presented a Monte-Carlo model for calculating the effect of initial 
and final state re-scattering in cold nuclear matter on the production
 and suppression 
of $\jpsi$ in heavy ion collisions. 
Although largely motivated by 
an earlier work from Qiu, Vary, and Zhang (QVZ)~\cite{Qiu:1998rz}, we obtain 
significantly different results at a given rescattering "strength" 
$\av{k_T^2}$ due to the use of 
{\em incoherent}
scatterings, which give considerable fluctuations in the
change of relative $c\cbar$ pair momentum as opposed to the 
overall shift 
expected intuitively based on coherent re-scattering studies of 
jet photoproduction and 
Drell-Yan\cite{Luo:1993ui,Benesh:1994,Fries:2003}.
Similar observations were previously discussed by 
Fujii~\cite{Fujii:2003fq} 
based on a three-dimensional Gaussian random walk model. 
Our analytic and numerical results indicate that, 
for a more realistic 2D random walk, 
momentum transfer fluctuations have an even stronger effect. 
These results highlight the importance of calculating
the $q^2$ modification for a $c\cbar$ pair
undergoing coherent multiple scattering in cold nuclear matter,
including the first nonvanishing contribution to fluctuations in 
$\Delta q^2$\cite{Fries:2003}.

In addition, we demonstrate a direct 
correlation between 
$\Delta\av{p_T^2} \equiv \av{p_T^2}_{A{+}A} 
- \av{p_T^2}_{p{+}p}$ 
and the $\jpsi$ suppression factor $R_{AA}$ in the incoherent multiple 
scattering model.  
We explore $\jpsi$ suppression and 
$\Delta\av{p_T^2}$ in $\sqrt{s_{NN}}=17.2$~GeV $Pb{+}Pb$ 
and $\sqrt{s_{NN}}=200$~GeV $Au{+}Au$ collisions as a function of centrality 
for a wide range of $\av{k_T^2}$ parameter values,
and find that
it impossible for this class of models to {\em simultaneously}
satisfy both suppression and $\av{p_T^2}$ constraints
from available 
nucleus-nucleus data at SPS or RHIC.
It would be interesting to compare these results with 
the suppression - $\av{p_T^2}$ 
correlation given by coherent multiple scatterings.

The importance of heavy quarkonia is emphasized by recent puzzling
observations regarding open heavy flavor in $Au{+}Au$ reactions 
at RHIC. Spectra~\cite{PHENIX_electron_pt,STAR_electron_pt}
and azimuthal anisotropy 
$v_2(p_T) \equiv \av{\cos 2\phi}_{p_T}$~\cite{PHENIX_electron_v2} 
of ``non-photonic'' single electrons indicate a striking modification of heavy
 quark
momentum distributions in medium.
We are extending our Monte-Carlo calculation for $\jpsi$ by
incorporating these $c\cbar$ pairs into the  parton cascade 
model~\cite{Molnar:2001ux} 
to thus include not only cold nuclear matter effects, but those of the hot
medium as well. Finally, we note that our framework could also be applied to 
study nuclear effects on other heavy-flavor mesons, such as $b\bar{b}$ states, 
and open charm and bottom.

\section{Acknowledgments}  
We acknowledge a careful reading of the manuscript by David Silvermyr.

This work is  supported by the Director, 
Office of Science, Office of High Energy and Nuclear Physics, 
Division of Nuclear Physics, of the U.S. Department of Energy 
under Grant No. DE-FG02-00ER41152 (AMG and JLN). DM thanks RIKEN, 
Brookhaven National Laboratory and
the US Department of Energy [DE-AC02-98CH10886] for providing facilities
essential for the completion of this work.


\begin{thebibliography}{99}
\bibitem{denes:foot2}
 The average $Q_0^2=q_z^2 - q_0^2$ 
 is largely independent of the $\jpsi$ rapidity $y$. However,
 the average $q_0^2$ grows with $|y|$.
\bibitem{denes:footnote3}
	This assumption is also made by studies 
	of radiative parton energy loss~\cite{BDMPS,GLV}.
\end{thebibliography}
\end{document}